\documentclass[12pt]{article}
\usepackage{graphicx}

\usepackage{multirow}
\usepackage{ctable}
\usepackage{booktabs}
\usepackage{array}
\usepackage{tabularx}
\usepackage{xcolor}
\usepackage{pstricks}
\DeclareMathAlphabet{\mathpzc}{OT1}{pzc}{m}{it}
\newcommand{\sigmaefftps}{\sigma_{\rm eff,TPS}}
\newcommand{\sigmaeffdps}{\sigma_{\rm eff,DPS}}

\newcommand\pubnumber{}
\newcommand\pubdate{\today}

\def\napoli{Institute of Nuclear
Physics, Polish Academy of Sciences, Radzikowskiego 152, PL-31-342 Krak{\'o}w, Poland}
\def\support{\footnote{This study was partially
supported by the Polish National Science Center grant
DEC-2014/15/B/ST2/02528 and by the Center for Innovation and
Transfer of Natural Sciences and Engineering Knowledge in
Rzesz{\'o}w.}}

\def\Title#1{\begin{center} {\Large #1 } \end{center}}
\def\Author#1{\begin{center}{ \sc #1} \end{center}}
\def\Address#1{\begin{center}{ \it #1} \end{center}}

\newcommand\pubblock{\rightline{\begin{tabular}{l} \pubnumber\\
         \pubdate  \end{tabular}}}
\newenvironment{Abstract}{\begin{quotation}  }{\end{quotation}}
\newenvironment{Presented}{\begin{quotation} \begin{center} 
             PRESENTED AT\end{center}\bigskip 
      \begin{center}\begin{large}}{\end{large}\end{center} \end{quotation}}

\def\beq{\begin{equation}}
\def\eeq#1{\label{#1}\end{equation}}
\def\eeqn{\end{equation}}

\def\beqa{\begin{eqnarray}}
\def\eeqa#1{\label{#1}\end{eqnarray}}
\def\eeqan{\end{eqnarray}}

\let\bar=\overbar

\def\L{{\cal L}}

\def\Dslash{\not{\hbox{\kern-4pt $D$}}}
\def\dslash{\not{\hbox{\kern-2pt $\del$}}}

\def\msb{{\bar{\ssstyle M \kern -1pt S}}}

\begin{document}
\begin{titlepage}
\pubblock

\vfill
\Title{Triple parton scattering effects in $D$-meson production\\ at the LHC}
\vfill
\Author{ Rafa{\l} Maciu{\l}a and Antoni Szczurek\support}
\Address{\napoli}
\vfill
\begin{Abstract}
We study triple-parton scattering effects in open charm production in proton-proton collisions at the LHC.
Predictions for one, two and three $c\bar c$ pairs production are given for $\sqrt{s}= 7$ TeV and $\sqrt{s}= 13$ TeV. Quite large cross sections, of the order of milibarns, for the triple-parton scattering mechanism are obtained. We suggest a measurement of three $D^{0}$ or three $\bar{D^{0}}$ mesons by the LHCb collaboration. The predicted visible cross sections are of the order of a few nanobarns. The counting rates including $D^{0} \to K^{-}\pi^{+}$ branching fractions are also given. We predict that at $\sqrt{s}= 13$ TeV a few thousands of events of triple-$D^{0}$ production can be observed by the LHCb collaboration.  
\end{Abstract}
\vfill
\begin{Presented}
The 17th Conference on Elastic and Diffractive Scattering, EDS Blois 2017,
26th - 30th June 2017, Prague, Czech Republic
\end{Presented}
\vfill
\end{titlepage}
\def\thefootnote{\fnsymbol{footnote}}
\setcounter{footnote}{0}

\section{Introduction}

The multi-parton interactions (MPI) got new impulse with the start
of the LHC operation \cite{Astalos:2015ivw,Proceedings:2016tff}. There are several ongoing studies of different
processes, so far mostly concentrated on phenomena of double-parton
scattering (DPS). Some time ago we have shown that charm production should
be one of the best reaction to study double-parton scattering effects \cite{Luszczak:2011zp}.
This was confirmed by the LHCb experimental data \cite{Aaij:2012dz} and
their subsequent interpretation \cite{Maciula:2013kd,vanHameren:2014ava,Maciula:2016wci}.

Very recently also triple parton scattering (TPS) was discussed in 
the context of multiple production of $c \bar c$ pairs \cite{dEnterria:2016ids}.
Inspiringly large cross sections were presented there. We followed this first analysis with application to triple $D$ meson production
and tried to answer the question whether the triple-parton scattering
could be seen in three $D^0$ or three $\bar D^0$ production at the LHCb experiment.

\section{Formalism}

The cross section for TPS in a general form \cite{Snigirev:2016uaq} can be written as follows:
\begin{eqnarray} 
\label{hardAB}
\sigma^{\rm TPS}_{pp \to c \bar c c \bar c c \bar c }  &=& \left(\frac{1}{3!}\right) \int \; \Gamma^{ggg}_{p}(x_1, x_2, x_3; {\vec b_1},{\vec b_2}, {\vec b_3}; \mu^2_1, \mu^2_2, \mu^2_3)\nonumber \\
& & \quad \times \quad \hat{\sigma}_{c\bar c}^{gg}(x_1, x'_1,\mu^2_1) \; \hat{\sigma}_{c\bar c}^{gg}(x_2, x'_2,\mu^2_2) \; \hat{\sigma}_{c\bar c}^{gg}(x_3, x'_3,\mu^2_3)\nonumber\\
& & \quad \times \quad \Gamma^{ggg}_{p}(x'_1, x'_2, x'_3; {\vec b_1} - {\vec b},{\vec b_2} - {\vec b},{\vec b_3} - {\vec b}; \mu^2_1, \mu^2_2, \mu^2_3)\nonumber\\
& & \quad \times \quad dx_1 \; dx_2 \; dx_3 \; dx'_1 \; dx'_2 \; dx'_3 \; d^2b_1 \; d^2b_2 \; d^2b_3 \; d^2b,
\end{eqnarray}
where $\hat{\sigma}_{c\bar c}^{gg}(x_i, x'_i,\mu^2_i)$ are the partonic cross sections for $gg\to c\bar c$ mechanism, $x_{i}$, $x'_{i}$ are the longitudinal momentum fractions, $\mu_{i}$ are the renormalization/factorization scales and $\frac{1}{3!}$ is the combinatorial factor relevant for the case of the three identical final states. 
The above TPS hadronic cross section is expressed in terms of the so-called triple-gluon distribution functions
$\Gamma^{ggg}_{p}(x_1, x_2, x_3; {\vec b_1},{\vec b_2}, {\vec b_3}; \mu^2_1, \mu^2_2, \mu^2_3)$.

The triple parton distribution functions (triple PDFs) shall account for all possible correlations between the partons. The MPI theory in this general form is well established (see \textit{e.g.} Ref.~\cite{Diehl:2011yj}) but not yet fully available for phenomenological investigations. The double PDFs in the case of DPS are under intense theoretical studies but their adoption to real process calculations is still limited. On the other hand, the objects like triple PDFs for TPS
were discussed so far only in Ref.~\cite{Snigirev:2016uaq}. 

As a consequence, in practice one usually follows the factorized Ansatz, where the correlations between partons are neglected and longitudinal and transverse degrees of freedom are separated. According to this approach the formula for inclusive TPS cross section (Eq.~(\ref{hardAB})) can be simplified to the pocket form \cite{dEnterria:2016ids}:
\begin{equation} 
\label{doubleAB}
\sigma_{pp \to c \bar c c \bar c c \bar c }^{\rm TPS} =  \left(\frac{1}{3!}\right)\, \frac{\sigma_{pp \to c \bar c}^{\rm SPS} \cdot
\sigma_{pp \to c \bar c}^{\rm SPS} \cdot \sigma_{pp \to c \bar c}^{\rm SPS}}{\sigmaefftps^2},
\end{equation} 
where the triple-parton scattering normalization factor $\sigmaefftps$ contains only information about proton transverse profile.

In principle, the DPS normalization factor $\sigmaeffdps$ was extracted experimentally from several Tevatron and LHC measurements (see \textit{e.g.} Refs.~\cite{Astalos:2015ivw,Proceedings:2016tff} and references therein) and its world average value is $\sigmaeffdps \simeq 15 \pm 5$ mb. Such experimental inputs are not available for $\sigmaefftps$. However, as was shown in Ref.~\cite{dEnterria:2016ids} for proton-proton collisions, the latter quantity can be expressed in terms of their more known DPS counterpart: 
\begin{equation}
\label{eq:TPS_DPS_factor}
\sigmaefftps = k\times\sigmaeffdps, \; {\rm with}\;\; k = 0.82\pm 0.11\,.
\end{equation}
The relation is valid for different (typical) parton transverse profiles of proton.
In the numerical calculations below we take $\sigmaeffdps = 21$ mb which corresponds to the average value extracted by the LHCb experiment only from the double charm data \cite{Aaij:2012dz}. This input gives us the value of $\sigmaefftps \simeq 17$ mb. 

In this paper, each of the single-parton scattering cross sections $\sigma_{pp \to c \bar c}^{\rm SPS}$ in Eq.~(\ref{doubleAB}) is calculated in the $k_{T}$-factorization approach \cite{kTfactorization}.
It was shown in Refs.~\cite{Maciula:2013wg,Maciula:2013kd,vanHameren:2014ava,Maciula:2016wci} that within this approach one can get a very good description of the LHCb single and double charm data.
In this approach the differential SPS cross section for inclusive single $c\bar c$ pair production can be written as:             
\begin{eqnarray}
\frac{d \sigma_{pp \to c \bar c}^{\rm SPS}}{d y_1 d y_2 d^2 p_{1,t} d^2 p_{2,t}}
&& = \frac{1}{16 \pi^2 {(x_1 x_2 S)}^2} \int \frac{d^2 k_{1t}}{\pi} \frac{d^2 k_{2t}}{\pi}
\overline{|{\cal M}_{g* g* \rightarrow c \bar{c}}|^2} \nonumber \\
&& \times \;\; \delta^2 \left( \vec{k}_{1t} + \vec{k}_{2t} - \vec{p}_{1t} - \vec{p}_{2t}
\right)
{\cal F}_{g}(x_1,k_{1t}^2,\mu^2) {\cal F}_{g}(x_2,k_{2t}^2,\mu^2),
\label{ccbar_kt_factorization}
\end{eqnarray}
where ${\cal M}_{g^* g^* \rightarrow c \bar{c}}$ is the well-known gauge-invariant off-shell matrix element for $g^* g^* \to c \bar c$ partonic subprocess and ${\cal F}_{g}(x_i,k_{it}^2,\mu^2)$ are the so-called unintegrated (transverse momentum dependent) gluon PDFs (uPDFs). Here we use the Kimber-Martin-Ryskin (KMR) uPDFs \cite{KMR}. In the perturbative part of the calculations, for the central predictions, we set both the renormalization and factorization scales equal to the averaged transverse mass $\mu^{2} = \frac{m^{2}_{i,t}+m^{2}_{j,t}}{2}$, where $m_{i,t} = \sqrt{p^{2}_{i,t} + m^{2}_{c}}$ and use the charm quark mass $m_{c} = 1.5$ GeV. 

The parton-level cross sections for triple charm quark (or charm antiquark) production are further corrected for the $c \to D$ (or $\bar c \to \bar{D}$) hadronization effects via the following procedure:   
\begin{equation}
\frac{d\sigma_{pp \to D D D }^{\rm TPS}}{d\xi^{D}}
 \approx
\int \frac{D_{c \to D}(z_{1})}{z_{1}}\cdot \frac{D_{c \to D}(z_{2})}{z_{2}} \cdot \frac{D_{c \to D}(z_{3})}{z_{3}}\cdot
\frac{d\sigma_{pp \to c c c }^{\rm TPS}}{d\xi^{c}} d z_{1} d z_{2} d z_{3} \; ,
\end{equation}
where $d\xi^{a}$ stand for $d y^{a}_1 d y^{a}_2 d y^{a}_3 d^2 p^{a}_{1,t} d^2 p^{a}_{2,t} d^2 p^{a}_{3,t}$ taking $a = c$ quark or $D$ meson and
$p^{c}_{i,t} = \frac{p_{i,t}^{D}}{z_{i}}$ with meson momentum fractions  $z_{i}\in (0,1)$. In the numerical calculations here we use the commonly used in the literature scale-independent Peterson \cite{Peterson:1982ak} fragmentation function $D_{c \to D}(z)$ with the parameter $\varepsilon_{c} = 0.05$. In the last step the obtained cross sections for triple-meson production are normalized with the corresponding
fragmentation fraction $\mathrm{BR}(c \to D^{0}) = 0.565$ \cite{Lohrmann:2011np}.

\section{Numerical results}

\begin{figure}[!h]
\begin{minipage}{0.47\textwidth}
 \centerline{\includegraphics[width=1.0\textwidth]{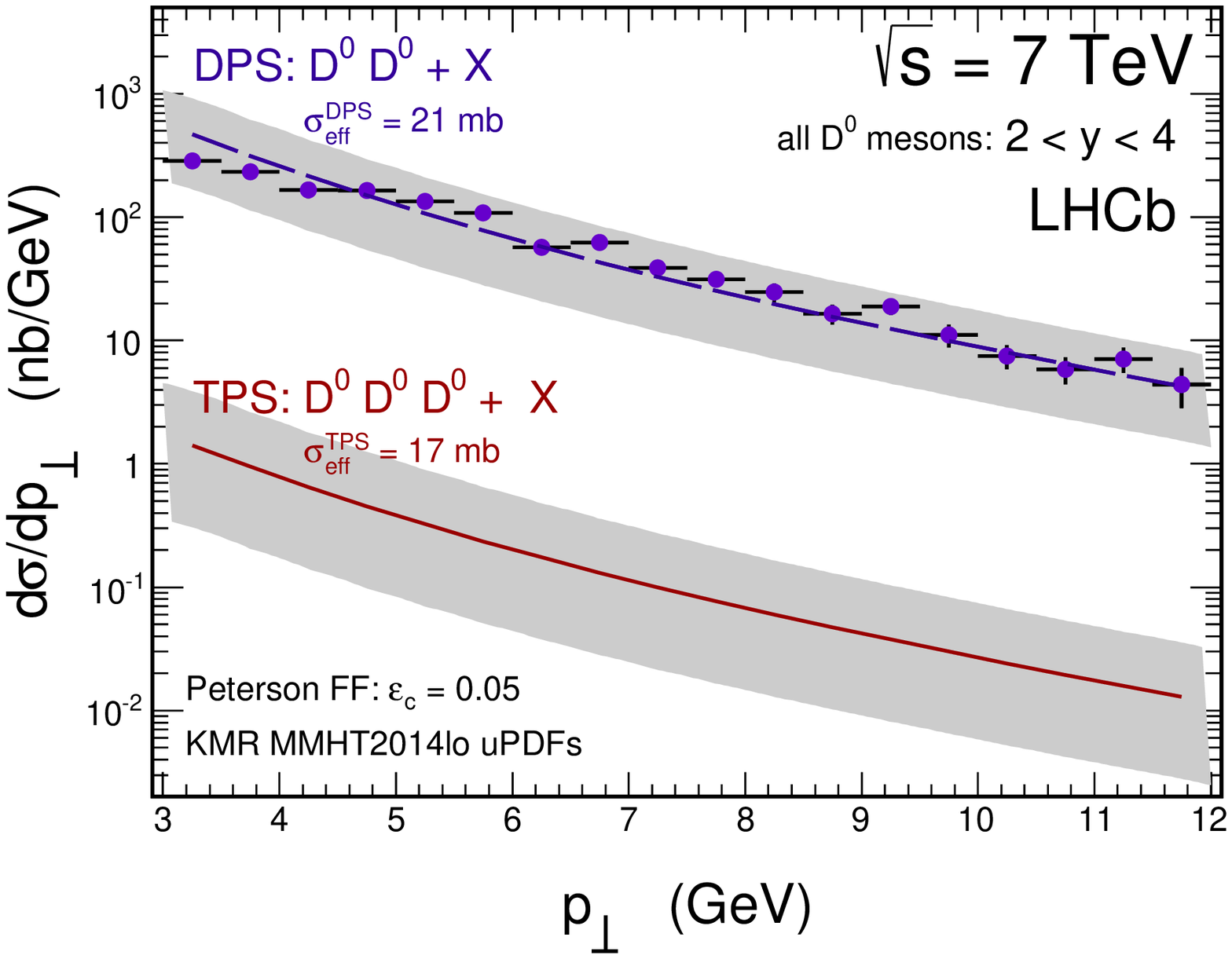}}
\end{minipage}
\hspace{0.5cm}
\begin{minipage}{0.47\textwidth}
 \centerline{\includegraphics[width=1.0\textwidth]{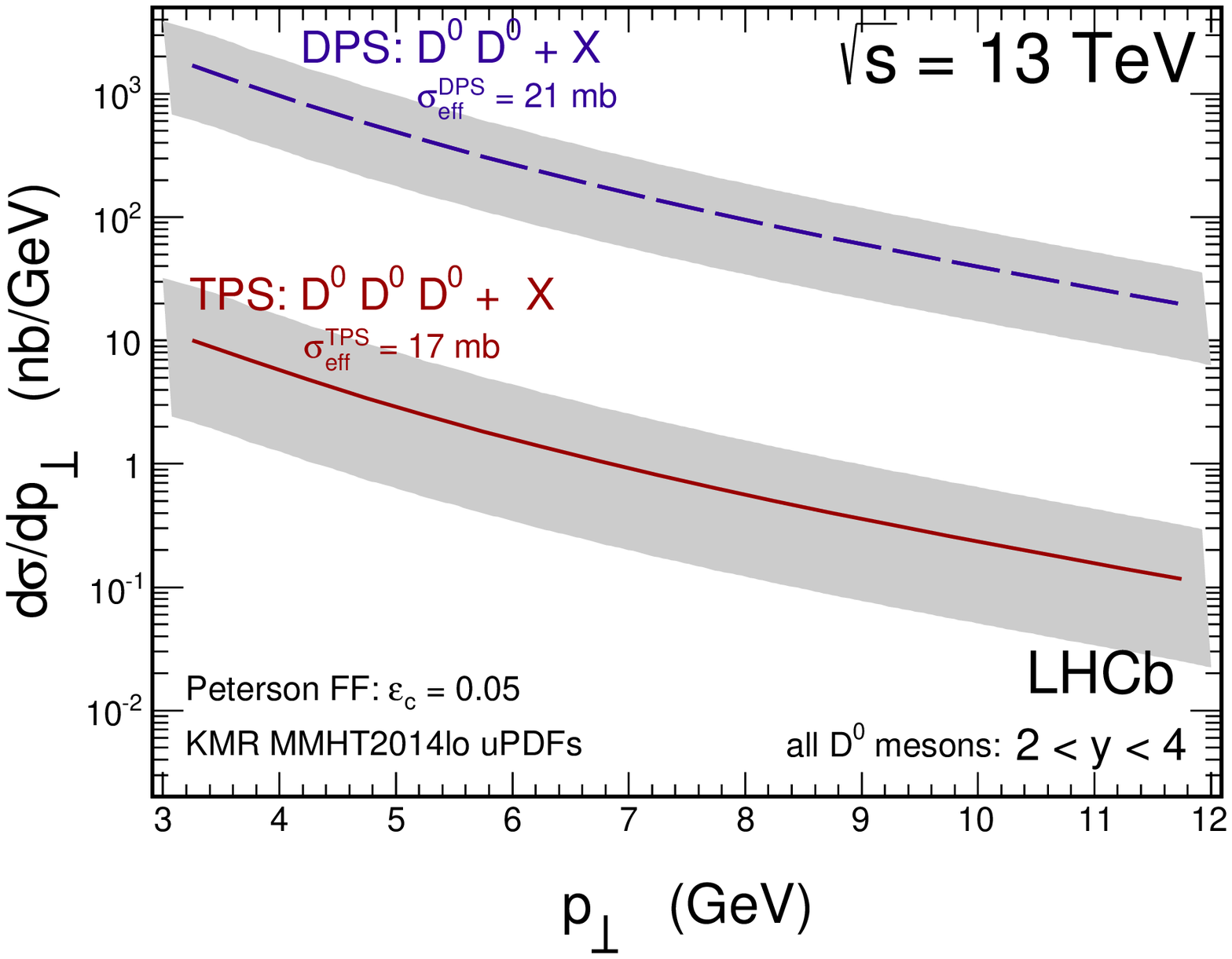}}
\end{minipage}
   \caption{
\small Transverse momentum distributions of one of the measured $D^{0}$
mesons for double-$D^{0}$ (upper long-dashed lines) and triple-$D^{0}$
(lower solid lines) production for $\sqrt{s} = 7$ (left panel) and $13$
TeV (right panel). Details are specified in the figure. The data points
for double-$D^{0}$ production in the left panel are taken from
Ref.~\cite{Aaij:2012dz}. The lower and upper limits of uncertainty bands were obtained from the lower and upper limits fors single $D^{0}$ ($\bar{D^{0}}$) production.
 }
 \label{fig:double-triple-D0}
\end{figure}

In Fig.~\ref{fig:double-triple-D0} we show transverse momentum distribution of one 
of the two or one of the three $D^0$ mesons (all measured by the LHCb detector).
In the used multiple parton scattering formalism the distributions
for single, double and triple production have the same shape/slope
and differ only by normalization. The distributions
for triple $D^0$ production are about two orders of magnitude smaller
than for double $D^0$ production, consistent with Table \ref{tab:total-cross-sections-D0}.
In the left panel, for $\sqrt{s}$ = 7 TeV, we also show for reference
the LHCb experimental data \cite{Aaij:2012dz}. We get a good agreement with the LHCb data within the uncertainty bands.
The uncertainties for the two and three meson case are propagated from the standard single-parton scattering (SPS) pQCD calculation uncertainties.
The chosen uncertainty procedure leads us to the conclusion that our result for TPS is known within a factor $\approx 3$.

In Table \ref{tab:total-cross-sections-D0} we show our predicted cross
sections for two and three mesons within the fiducial volume
of the LHCb detector. The predicted value at $\sqrt{s}$ = 7 TeV for $D^{0}D^{0} + \bar{D^{0}}\bar{D^{0}}$ final state is
consistent with the measured one: $\sigma_{LHCb} =  690 \pm 40 \pm 70$ (see Table 12 in Ref.~\cite{Aaij:2012dz}).

\begin{table}[tb]%
\caption{The integrated cross sections for double and triple $D^{0}$ meson production (in nb) within the LHCb acceptance: $2 < y_{D^{0}} < 4$ and $3 < p_{T}^{D^{0}} < 12$ GeV, calculated in the $k_{T}$-factorization approach. The numbers include also the charge conjugate states.}
\label{tab:total-cross-sections-D0}
\centering %
\newcolumntype{Z}{>{\centering\arraybackslash}X}
\begin{tabularx}{1.0\linewidth}{Z Z Z}
\toprule[0.1em] %

Final state  & $\sqrt{s} = 7$ TeV  & $\sqrt{s} = 13$ TeV       \\ [-0.2ex]

\bottomrule[0.1em]

DPS: $\sigma(D^{0}D^{0} + X)$  & 784.74 $\;\;\;^{\mathrm{max:\;} 1538.09}_{\mathrm{min:\;} 282.51}$  & 2992.91 $\;\;\;^{\mathrm{max:\;} 5866.10}_{\mathrm{min:\;} 1077.45}$ \\  [-0.2ex]
TPS: $\sigma(D^{0}D^{0}D^{0} + X)$  & 2.38 $\;\;\;^{\mathrm{max:\;} 6.53}_{\mathrm{min:\;} 0.51}$  & 17.71 $\;\;\;^{\mathrm{max:\;} 48.59}_{\mathrm{min:\;} 3.83}$ \\  [-0.2ex]

\bottomrule[0.1em]

\end{tabularx}
\end{table}

Finally in Table \ref{tab:events} we show the number of counts
for different realistic values of the integrated luminosity for the LHCb experiment. The predicted numbers of events for DPS double- and TPS triple-$D^{0}$
production correspond to the central predictions for cross sections from Table~\ref{tab:total-cross-sections-D0}. Here we have included
in addition the relevant decay branching fraction $\mathrm{BR}(D^{0} \to K^{-} \pi^{+}) = 0.0393$ \cite{Olive:2016xmw}.
In the case of triple $D^0$ production we predicted about 100 counts at $\sqrt{s}= 7$ TeV and a few thousands of counts at $\sqrt{s}= 13$ TeV
for realistic integrated luminosities. We hope the LHCb collaboration will be able to verify our
predictions soon. More details of the analysis can be found in our original paper \cite{Maciula:2017meb}.

\begin{table}[tb]%
\caption{Number of events for different values of the feasible integrated luminosity in the LHCb experiment for the central predictions of cross sections from Table~\ref{tab:total-cross-sections-D0}. The branching fractions for $D^{0} \to K^{-}\pi^{+} (\bar{D^{0}} \to K^{+}\pi^{-})$ are included here.}
\label{tab:events}
\centering %
\newcolumntype{Z}{>{\centering\arraybackslash}X}
\begin{tabularx}{1.0\linewidth}{Z Z Z Z}
\toprule[0.1em] %

$\sqrt{s}$ & Integrated Luminosity & DPS ($D^{0}D^{0}$) & TPS ($D^{0}D^{0}D^{0}$) \\

\bottomrule[0.1em]

 \multirow{2}{1.2cm}{$7$ TeV} &  355 pb$^{-1}$   &  $0.43 \times 10^{6}$   & 51    \\
                             &  1106 pb$^{-1}$  &  $1.34 \times 10^{6}$    & 159     \\
\hline
 \multirow{2}{1.2cm}{$13$ TeV} &  1665 pb$^{-1}$   &  $7.70 \times 10^{6}$ & 1789    \\
                             &  5000 pb$^{-1}$  &  $23.11 \times 10^{6}$   & 5374     \\                             


\bottomrule[0.1em]

\end{tabularx}
\end{table}

\end{document}